\documentclass[10pt,conference]{IEEEtran}

\usepackage{enumitem}
\setlist[itemize]{topsep=2pt, itemsep=1pt, parsep=0pt, partopsep=0pt}

\usepackage{amsmath,amssymb,amsfonts}
\usepackage{url}

\usepackage{textcomp}
\usepackage{xcolor}
\usepackage{graphicx}
\usepackage{url}
\usepackage{subfig}
\usepackage{subcaption}
\usepackage{placeins}

\usepackage{caption}
\usepackage{multirow}
\usepackage{array}
\usepackage{float}
\usepackage{hyperref}
\usepackage{pifont}
\usepackage{listings}
\usepackage{balance}
\usepackage{cmap}  
\usepackage{threeparttable}
\usepackage{colortbl}
\usepackage{booktabs}
\usepackage{tabularx}
\usepackage{adjustbox}

\definecolor{grayboxcolor}{HTML}{f2f2f2}

\title{Prompt Structure Redistributes, Not Reduces: An Empirical Analysis of Security-Weaknesses in LLM-Generated Python Code}

\author{
\IEEEauthorblockN{
Maitreyee Das Urmi\IEEEauthorrefmark{1},
Jessica Pourleyli\IEEEauthorrefmark{1},
Fabio Santos\IEEEauthorrefmark{2},
Glaucia Melo\IEEEauthorrefmark{1}
}

\IEEEauthorblockA{
\IEEEauthorrefmark{1}
\textit{Toronto Metropolitan University},
Toronto, Canada \\
\{maitreyee.urmi, jessica.pourleyli, glaucia\}@torontomu.ca
}

\IEEEauthorblockA{
\IEEEauthorrefmark{2}
\textit{Colorado State University},
Fort Collins, USA \\
fabio.deabreusantos@colostate.edu
}
}

\begin{document}

\maketitle

\begin{abstract}
Large Language Models (LLMs) increasingly generate code from natural-language prompts, making prompt engineering a key mechanism for shaping the security of generated software. Structured and security-oriented prompts are widely used to encourage safer code, yet their effects extend beyond whether detected weaknesses are simply present or absent. Using 424 security-sensitive Python tasks, we generate solutions with GPT-4o and LLaMA 3.1-8B under five prompt variants that progressively add structural and security guidance, and evaluate them with Bandit and CodeQL along two axes: generation compliance and security weakness prevalence, severity, and CWE distributions. Structured prompting substantially reduces refusals (e.g., GPT-4o invalid outputs drop from 338 of 424 to 37-52), enabling large-scale analysis, but security-oriented refinements do not consistently reduce overall weakness prevalence. For GPT-4o, stronger prompts primarily redistribute risk: high-severity findings fall (20.8\%→13.6\%) while low-severity findings rise (32\%→43.5\%); LLaMA shows weaker, less consistent shifts. We also observe security-driven semantic drift, where stricter prompts silently remove or rewrite explicitly requested unsafe constructs. Overall, prompt structure improves compliance but is an unreliable substitute for robust security controls in LLM-assisted development.

\end{abstract}


\maketitle

\section{Introduction}

Large Language Models (LLMs) have rapidly transformed modern software engineering by enabling developers to generate, refactor, and analyze code via natural-language instructions. However, as LLMs are increasingly integrated into production pipelines, their outputs are not free from risk. Studies have shown that these models often produce inconsistent or error-prone code, requiring manual verification and iteration to ensure correctness~\cite{murr2023specificity,dellaporta2025patterns}. More critically, when prompts are vague or underspecified, LLMs may introduce unsafe imports, weak authentication logic, or injection vulnerabilities, raising important questions about how prompt design impacts code security.

Prompt engineering has emerged as a key factor influencing model performance~\cite{khojah2025impact}. Structured prompts that incorporate explicit context, constraints, or role definitions have been shown to improve code quality, readability, and maintainability, particularly when combined with persona-driven prompting or semantic patterning. However, existing studies largely emphasize functional correctness rather than the security properties of generated code. At the same time, research on prompt robustness demonstrates that LLM outputs are highly sensitive to prompt phrasing: even minor perturbations can yield unstable or incorrect responses~\cite{zhu2024promptrobust}, and adversarially crafted prompts can introduce backdoors into prompt-based models~\cite{yao2023poisonprompt}. Together, previous research reveals both the power and fragility of prompt design, suggesting an overlooked opportunity: if structured prompting can improve correctness and consistency, it may also shape how security weaknesses manifest during code generation, influencing their distribution even when overall prevalence remains largely unchanged.

To address that, rather than treating prompts solely as interfaces for improving accuracy or efficiency, we investigate them as potential mechanisms for influencing the security characteristics of generated code and find that the dominant effect is not reduction but redistribution. We inquire into the prevailing aggregate-reduction narrative in secure code generation research: prior work often reports decreases in static-analysis findings under security-oriented prompting and concludes that prompting improves security. To test this hypothesis, we design an empirical study that systematically assesses prompt structures for identical programming tasks, generates code snippets using representative proprietary and open-weight LLMs, and evaluates them using \textit{Bandit}\footnote{https://bandit.readthedocs.io/en/latest/index.html} and \textit{CodeQL\footnote{https://codeql.github.com/}}, static analysis tools for security-weakness pattern detection. Our goal is to characterize how prompt structure influences both the likelihood that models generate code for security-sensitive tasks and the severity of security weaknesses in the resulting outputs, thereby establishing an evidence-based understanding of what prompt engineering can and cannot achieve for secure code generation.

This work addresses the following research questions, each motivated by the need to understand how prompt structure shapes the prevalence, severity, and distribution of security weaknesses in LLM-generated code.

\textbf{RQ1: How does prompt structure affect code-generation compliance for security-sensitive tasks?}\\
Prior work shows that structured prompts can improve correctness and maintainability~\cite{murr2023specificity,dellaporta2025patterns} and that prompt formulation can meaningfully affect model behaviour~\cite{khojah2025impact, zhu2024promptrobust}. We measure model compliance by distinguishing between valid, executable Python code and refusal/non-code outputs, where refusals are plain-text responses in which the model declines to generate code for security-sensitive programming tasks that could lead to insecure implementations.

\textbf{RQ2: Conditioned on valid code, how does prompt structure affect security risk in generated outputs?}\\
Conditioned on valid code, we evaluate whether a security-oriented prompt structure changes the prevalence, severity, and CWE distributions of security weaknesses in the resulting code.

Our findings highlight important limitations of prompt-level security guidance and motivate the need for complementary mitigation mechanisms. In summary, this study contributes:
\begin{enumerate}
    \item \textbf{A two-stage evaluation framework} that separates code-generation compliance from downstream security analysis, revealing that prompt refinements can substantially affect the population of analyzable outputs and thereby confound weakness comparisons.
    \item \textbf{Empirical evidence that structured prompting is primarily a compliance lever}, dramatically reducing refusals for GPT-4o but not consistently for LLaMA 3.1-8B.
    \item \textbf{A semantic-drift audit} as a new analytical lens, showing that security-guided prompts silently alter explicitly requested constructs in ways that satisfy static analyzers while potentially violating functional requirements.
    \item \textbf{A characterization of prompt-level security effects}, showing that structural, security, framework, and adversarial guidance influence compliance and security weaknesses in distinct ways rather than producing uniform improvement.
    \item \textbf{A publicly available replication package} containing all prompts, generated outputs, and static-analysis results across both models and five prompt variants, enabling independent verification and reuse~\cite{anonymous2026empirical}.
\end{enumerate}


\section{Related Work}
This work intersects three lines of research: the quality and security of LLM-generated code, prompt engineering for software engineering tasks, and security-aware code generation. Across these areas, security is most often assessed through aggregate weakness counts. As a result, prior work says little about how interventions shift risk across severity levels rather than reducing it outright, the question our study takes up.

\subsection{LLM-Generated Code Quality and Security Weaknesses}

A growing body of work evaluates the quality and security of LLM-generated code. Early large-scale studies show that code models can produce syntactically correct and functionally plausible programs, though performance degrades with task complexity~\cite{chen2021evaluating}. Subsequent security-focused analyses reveal that generated code frequently contains unsafe patterns and vulnerabilities~\cite{pearce2025asleep}, even when functional tests pass~\cite{mohsin2024can}. Empirical evaluations report substantial vulnerability rates in LLM-generated outputs~\cite{pearce2025asleep}, motivating the use of static analysis tools and vulnerability-focused benchmarks such as CodeLMSec~\cite{hajipour2024codelmsec,li2024iris}. For example, CodeLMSec demonstrates that security weaknesses persist even when models exhibit strong functional performance~\cite{hajipour2024codelmsec}, and analyses of GitHub Copilot report that roughly 40\% of suggestions in security-sensitive contexts contain vulnerabilities~\cite{pearce2025asleep}.

These findings motivate fine-grained analyses of vulnerability characteristics rather than binary or count-based metrics, a direction we pursue at the level of severity rather than prevalence.

\subsection{Prompt Engineering for Software Engineering Tasks}

Prompt engineering plays a central role in shaping LLM behaviour across software engineering tasks, including code generation, repair, and testing. Careful prompt formulation improves functional correctness and specification adherence~\cite{chen2021evaluating}, while structured strategies such as explicit instructions and constraint-based prompting enhance consistency and reliability~\cite{TonyPromptingTechniques2025}. Automated prompt refinement techniques, including prompt linting and repair, further improve output quality and mitigate bias or vulnerability issues~\cite{promptlinting2025}.

Beyond correctness, prompt design influences code quality attributes such as maintainability and reliability~\cite{maintainability2025}. In practical settings such as CRUD development, structured prompts improve static quality metrics and reduce certain defect classes, though security misconfigurations often persist across models and prompt variants~\cite{crudprompt2025}. More recently, prompt engineering has been applied to security-sensitive tasks, including vulnerability mitigation and responsible code generation~\cite{promptlinting2025,benchmarksecureprompt2025,homoliak2024enhancing}. 

While these studies show that carefully designed prompts can reduce detected vulnerabilities, sometimes targeting specific vulnerability types~\cite{benchmarksecureprompt2025}, they typically rely on aggregate reduction metrics and do not examine how security guidance affects generation compliance or the composition of remaining weaknesses.

\subsection{Security-Aware and Responsible Code Generation}

Recent work proposes security-aware prompting and optimization techniques to reduce security weaknesses in LLM-generated code, including explicit security instructions, iterative self-repair, and static-analysis feedback loops~\cite{promsec2024,guidingai2025,staticfeedback2025}. Empirical results generally report reductions in detected security weaknesses relative to baseline prompting. Some systems formalize secure code generation as an optimization problem balancing correctness and security objectives~\cite{promsec2024}, while others explore incremental or multi-stage prompting to improve adherence to secure coding practices~\cite{liu2024solitary}. However, these approaches typically evaluate effectiveness based on overall (total) reductions in detected weaknesses. Prior critiques argue that separating security from functionality and relying on a single static analysis tool can produce incomplete or misleading assessments~\cite{rethinkingsecureeval2025}. 

Despite this concern, existing studies primarily evaluate prompt-based security interventions using aggregate vulnerability counts. Consequently, little is known about whether apparent security improvements reflect true reductions in risk or shifts in security issue severity and composition. Our study addresses this gap by analyzing severity distributions rather than counts alone, showing that prompt-level interventions can reallocate risk across severity levels and introduce trade-offs that binary metrics fail to capture.

\section{Methodology}

\begin{figure}[htbp]
\centering
\includegraphics[
    width=0.5\textwidth
]{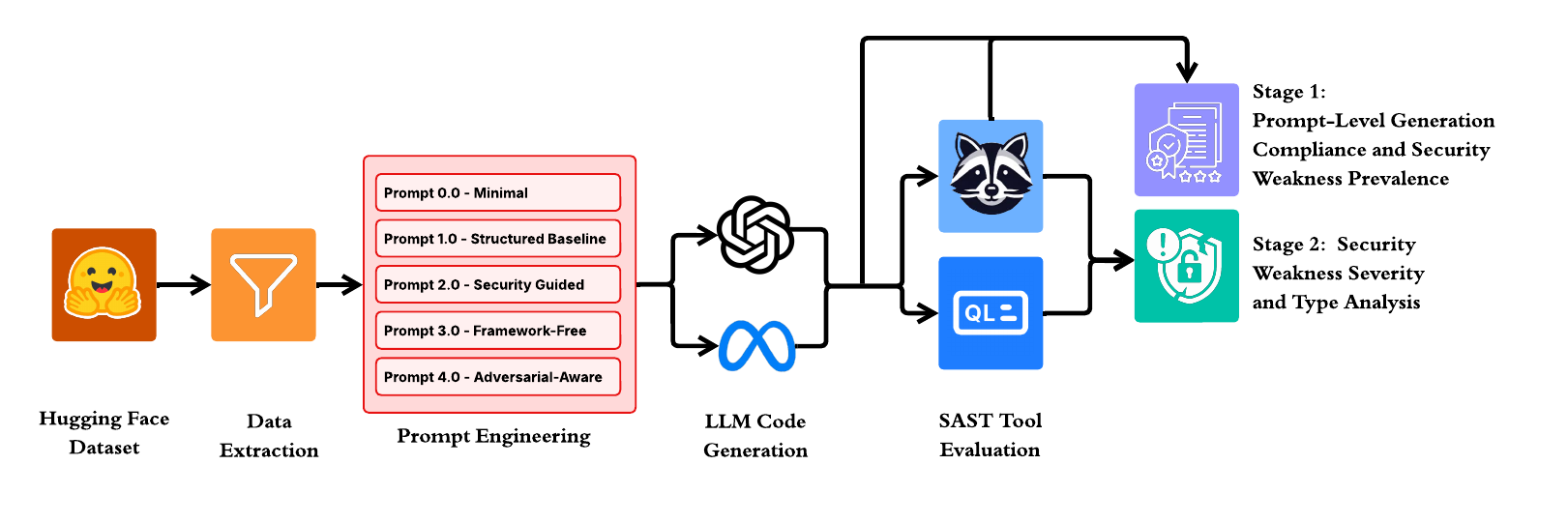}
\caption{Overview of the experimental pipeline.
}
\label{fig:method}
\end{figure}

To investigate how structured prompts influence both code-generation success and the severity of security weaknesses in LLM-generated code, we designed an empirical study that systematically varies prompt structures, generates code using representative proprietary and open-weight language models, and evaluates the resulting code via static analysis. Our methodology follows a three-stage process: (1) prompt design, (2) code generation and collection, and (3) security weakness assessment using Bandit and CodeQL. 
Figure~\ref{fig:method} provides an overview of our experimental pipeline.

\subsection{Experimental Infrastructure}
All experiments were conducted in a controlled local environment to ensure reproducibility and consistency across models, prompts and analysis tools. We evaluated two large language models: GPT-4o and LLaMA 3.1-8B. These models were selected to capture both a high-capacity proprietary model and a widely used open-weight model, reflecting common deployment scenarios in practice. We evaluate GPT-4o and LLaMA 3.1-8B to capture the two dominant deployment paradigms: a high-capacity, proprietary model accessed via a commercial API, and a widely used, open-weight model enabling local, reproducible execution. This lets us assess whether prompt-driven security effects are model-class-specific or generalizable. 

All code generation and analysis were performed using Python 3.11.5 within a Python virtual environment compiled with Clang 2.23.3. Security-relevant issues detection was conducted using Bandit and CodeQL, two widely adopted static analysis tools for identifying security weaknesses in Python programs. Bandit provides rule-based detection with LOW, MEDIUM, and HIGH severity annotations, while CodeQL enables semantic, query-driven vulnerability discovery, offering complementary coverage of security issues. Experiments were executed on a macOS system (arm64, Apple Silicon) equipped with an Apple M1 CPU and 8GB RAM. All computations were performed solely on the CPU. Using a consistent hardware and software environment reduces potential confounding factors, helping isolate the effects of prompt structure on the generated outputs. In addition to security issue findings, we record whether each prompt–model interaction yields executable Python code or results in a refusal, enabling analysis of prompt effects on code-generation success.

\subsection{Dataset Selection and Preparation}

We source security-sensitive Python coding tasks from the publicly available Cybernative.ai Code Vulnerability and Security Dataset \cite{cybernative_code_vuln_security_dpo}. The dataset was filtered to retain only Python examples, yielding 424 rows. These programming questions are embedded directly in the user prompt as input, instructing the model to produce a secure and valid Python solution to the given task. 

We refer to these tasks as \emph{security-sensitive} because they are built around vulnerability-prone programming scenarios rather than general-purpose algorithmic problems. Many involve operations such as command execution, file handling, deserialization, cryptographic functionality, or input validation, creating opportunities for insecure implementations that can be assessed through static analysis.

\subsection{Prompt Design}

Following prior work demonstrating that structured prompting improves LLM output quality and reliability \cite{zengPromptTemplate2025}, we design a set of related prompt templates that vary in system structure while maintaining a consistent output format. Each interaction consists of a system and a user prompt. 

\subsubsection{User Prompt (Invariant)}
The user prompt is held constant across all experiments:

\begin{quote}
Write a complete and correct code solution to the following problem: 

\{insecure\_coding\_question\}
\end{quote}

\subsubsection{Base Prompt Specifications}
 
\textbf{Persona / Role Definition.} All prompt variants assign the model the role of a \textit{Senior Software Engineer}, a common strategy for guiding domain-specific reasoning. Prior work shows that expert-role prompting can improve adherence to secure coding practices~\cite{TonyPromptingTechniques2025}. We adopt a software engineering persona, rather than a security engineering persona, to reflect zero-shot code-generation settings, emphasizing task completion over post hoc vulnerability analysis.

\textbf{Input Context and Output Expectations.} Each system prompt explicitly states that a coding task follows and specifies the expected output. Prompt template frameworks recommend clear input and output constraints to reduce ambiguity and improve compliance~\cite{zengPromptTemplate2025}. In pilot experiments, models often returned explanations rather than implementations; therefore, all variants explicitly require \textbf{code-only} outputs.

\textbf{Style, Quality, and Formatting Constraints.} Prompts enforce Python best practices (PEP-8), modular design, and explicit error handling to promote production-level quality and reduce code smells~\cite{fowler1999refactoring}. These constraints align with CSM principles of clarity, consistency, and modularity~\cite{KirkCodeStyleModel2025}. Strict output formatting enables automated extraction and evaluation.

\subsubsection{Prompt Variants and Strategy}
Before designing structured prompt variants, we began with a minimal baseline system prompt, referred to as \textbf{Prompt 0.0}, which simply instructed the model: \textit{"You are a senior software engineer. Generate complete and correct solutions for the given coding question."} 

\begin{figure}[htbp]
    \centering
    \includegraphics[
        width=\columnwidth
    ]{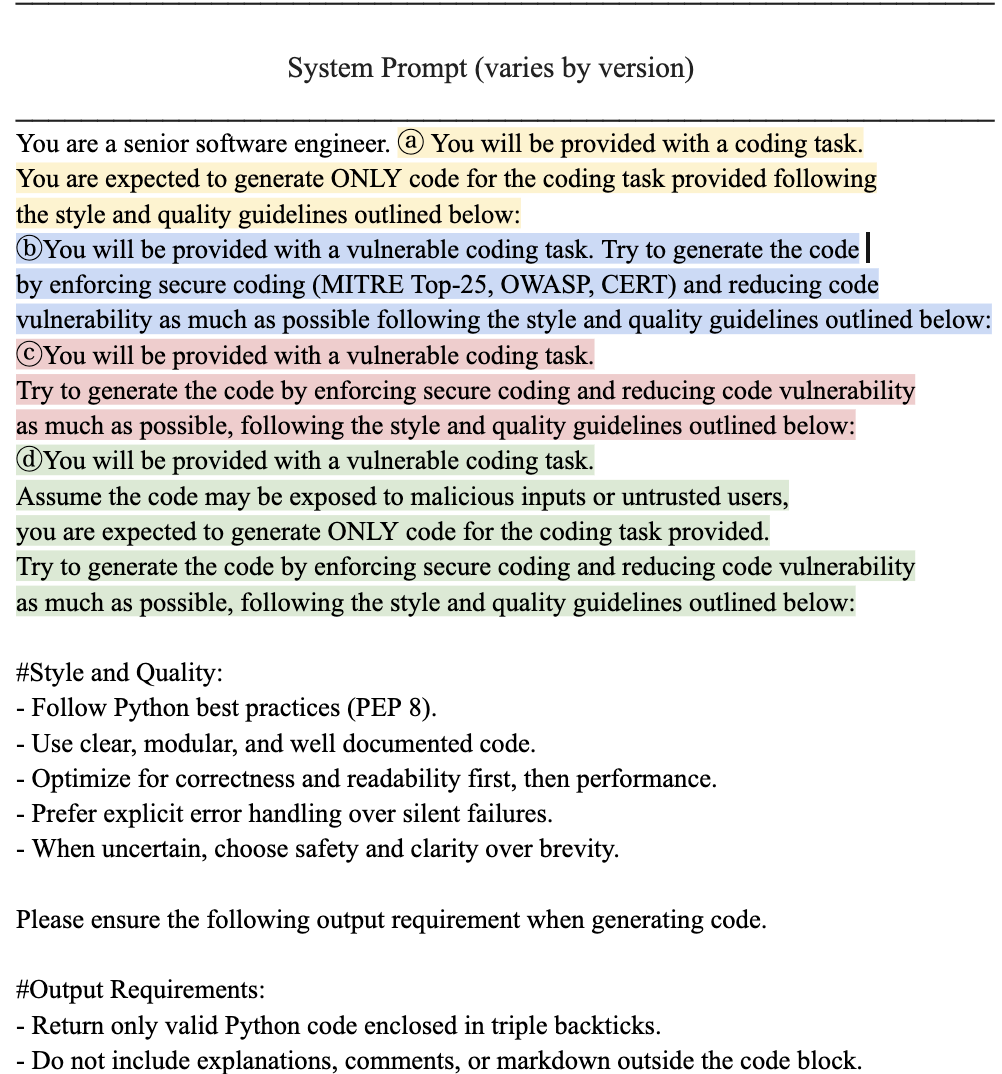}
    \caption{System prompt components for Prompts 1.0--4.0. Highlighted regions (\textcircled{a}--\textcircled{d}) indicate modifications introduced in each variant.}
    \label{fig:combined_prompts}
\end{figure}

Under minimal prompting (Prompt 0.0), both models frequently produced refusals or non-code outputs for security-sensitive tasks, motivating additional structural constraints to improve compliance. We therefore introduce four incremental prompt variants that modify structural guidance and security emphasis while preserving style and output format (Figure~\ref{fig:combined_prompts}).

\begin{itemize}
    \item \textbf{Prompt 1.0 (\textcircled{a}, yellow): Structured Baseline} 
Extends Prompt 0.0 by adding explicit output constraints, including an instruction to generate \textbf{only} code, along with persona definition, input context, and style and quality guidelines. This variant operationalizes established prompt template frameworks that emphasize contextual grounding and explicit output specifications \cite{zengPromptTemplate2025}. Prompt 1.0 primarily targets the reduction of refusal behaviour and serves as the structured baseline. It introduces the persona definition, input context, explicit output constraints, and style and quality requirements described above.

    \item \textbf{Prompt 2.0 (\textcircled{b}, blue): Security-Guided} 
Extends Prompt 1.0 by explicitly instructing the model to follow well-known secure coding guidelines (e.g., OWASP, CERT, MITRE Top-25) and to minimize code insecurities. Prior work shows that incorporating high-level security objectives into prompts can reduce vulnerabilities in LLM-generated code \cite{TonyPromptingTechniques2025}. This variant tests whether adding high-level security objectives influences security weakness severity beyond the compliance gains achieved by Prompt 1.0.

    \item \textbf{Prompt 3.0 (\textcircled{c}, red): Framework-Free Security Reinforcement} 
A modification of Prompt 2.0 that removes explicit references to security frameworks such as OWASP, CERT, and the MITRE Top 25 Most Dangerous Software Weaknesses while retaining the instruction to write secure code and minimize security-relevant issues. This variant isolates the effect of security framing itself from the effect of citing established security standards.

    \item \textbf{Prompt 4.0 (\textcircled{d}, green): Adversarial-Aware} 
Modifies Prompt 3.0 by introducing an explicit adversarial context in which the generated code may be exposed to malicious inputs or untrusted users. This variant evaluates whether emphasizing defensive programming under threat assumptions affects generation behaviour or reduces the severity of security issues.
\end{itemize}

The prompt variants were designed as controlled experimental conditions rather than an iterative prompt-optimization process. Each variant introduces a distinct form of guidance (structural constraints, security guidance, framework references, or adversarial context) to isolate its effect on model behaviour. We limited the study to five variants because the goal was to compare a small set of interpretable prompting strategies rather than exhaustively explore the prompt design space. 
We use zero-shot prompting for all variants to isolate the effect of prompt structure. Techniques such as few-shot prompting, chain-of-thought, and self-refinement can improve performance but introduce additional variation and rely on model self-correction, which may be unreliable and task-dependent~\cite{TonyPromptingTechniques2025}. Zero-shot prompting provides a cleaner setting for analyzing the success of generation and the severity of security weaknesses.

\subsubsection{Prompt Refinement Rationale and Termination Criterion}

Each prompt variant was motivated by a specific hypothesis: Prompt 0.0 established a baseline, 1.0 tested whether structural scaffolding reduces refusals, 2.0 tested explicit security objectives, 3.0 tested whether named frameworks (OWASP, MITRE) drive the effect or whether framing alone suffices, and 4.0 tested adversarial awareness. Refinement stopped at 5 because the design space was systematically sampled along a single conceptual axis: none → structure → security guidance → security without frameworks → adversarial awareness. 

\subsection{Code Generation and Data Collection }

For each Python programming question in the dataset, we generate a single candidate solution for each prompt variant and model, yielding a total of 424 tasks × 5 prompt variants × 2 models. Each generation is performed independently using zero-shot prompting.


\textbf{Model Access and Configuration.}
GPT-4o is accessed through the OpenAI API, and LLaMA 3.1-8B is accessed locally via Ollama using \texttt{Q4\_K\_M} quantization. For both models, we set the temperature to 0.7 and the top-$p$ value to 0.2. Prior work indicates that moderately higher temperatures can improve model exploration and task performance in code generation settings \cite{liu2024solitary}. Accordingly, we adopt a temperature of 0.7 to balance output diversity with stability, while a low top-p constraint constrains sampling to high-probability tokens.

\textbf{Generation Procedure.}
Each prompt-model task combination is executed once to avoid bias introduced by cherry-picking or best-of-\emph{n} sampling, and the raw model output is then stored. We record whether the output contains executable Python code or constitutes a refusal/non-code response. Outputs were evaluated using an automated audit script that attempted to parse each output using Python's \texttt{ast.parse()} and verified that successfully parsed files contained executable code rather than only standalone strings or docstrings. Outputs that failed automated checks were manually reviewed to distinguish syntax errors from refusals and other non-code responses. No syntax-error outputs were observed for GPT-4o; for LLaMA 3.1-8B, only 1, 2, 2, and 3 syntax-error outputs were identified under Prompts 1.0--4.0, respectively. Consequently, invalid-output counts were overwhelmingly attributable to refusals and other non-code responses. Refusal rates are calculated as the proportion of tasks that do not produce executable Python code. Outputs are written to individual Python files and assigned unique identifiers based on problem ID, model, and prompt version. This procedure enables analysis of both (1) code generation success rates and (2) downstream security weakness characteristics, while maintaining consistent experimental conditions across all prompt variants.

\subsection{Security Weakness Analysis with Bandit and CodeQL}

We analyze all generated Python files using two complementary static application security testing tools: Bandit and CodeQL. Bandit provides lightweight Python-specific detection of common insecure coding patterns and reports severity levels, making it well-suited for large-scale profiling of security weaknesses. CodeQL complements this capability through semantic and data-flow analysis, enabling the detection of security weaknesses that depend on how information propagates through a program. Using both tools reduces dependence on a single analyzer and broadens coverage of security issues by combining pattern-based and semantic analysis approaches.

For CWE-level analysis, we focus on the top 10 most frequent Common Weakness Enumeration (CWE) identifiers. CWE is a standardized software weakness taxonomy maintained by MITRE and widely used for security weakness classification and reporting~\cite{mitre_cwe}. Bandit and CodeQL report findings together with associated CWE identifiers, enabling analysis of security weakness distributions across prompt variants. 




\subsection{Analysis and Validation}
Our analysis proceeds in two stages, described next.

\textbf{Stage 1 (RQ1):  Prompt-Level Generation
Compliance and Security Weakness Prevalence}
We first examine all prompt variants (Prompt 0.0--4.0) to assess:
(i) the number of invalid or non-code outputs, and
(ii) the number of files containing security weaknesses.
This stage establishes whether structured prompts primarily affect generation compliance, security-relevant issue severity, or both. 

\textbf{Stage 2 (RQ2): Security Weakness Severity and Type Analysis.}
At this stage, security weakness analysis is performed only on outputs that contain executable Python code, as static-analysis tools cannot be meaningfully applied to refusals or other non-code responses. For each model--prompt pair, Bandit and CodeQL are applied to all valid code outputs generated under that condition. Because the number of valid outputs varies across model--prompt pairs, we report security issue prevalence, severity distributions, and CWE frequencies as percentages rather than raw counts to enable meaningful comparison across conditions. Due to the high rate of non-code responses under GPT-4o Prompt 0.0 (approximately 80\% of outputs), Prompt 0.0 was excluded from the security weakness-focused analysis, and only Prompts 1.0--4.0 were considered for GPT-4o in Stage 2.

\begin{itemize}
    \item Bandit severity distributions across prompts
    \item Common Weakness Enumeration (CWE) frequency distributions across prompts
\end{itemize}

We first conduct this analysis on GPT-4o outputs alone, given their substantially higher rate of valid code generation. We then repeat the same analysis on the combined GPT-4o and LLaMA 3.1-8B subset to examine whether observed trends generalize across model classes.

\textbf{Validation Considerations.}
To ensure internal validity, all variables other than system prompt structure are held constant: dataset, user prompt, model parameters, hardware, and analysis tools. Using a single generation per configuration avoids sampling bias and reflects realistic usage scenarios. While this limits variance estimation, it enables a clean isolation of prompt-structure effects.

\textbf{Task Alignment Verification.}
Because the dataset contains intentionally vulnerable coding tasks, many programs cannot be safely executed for dynamic validation. Instead, we conduct a lightweight relevance audit to verify that prompt refinements do not reduce security weakness counts simply by producing degenerate outputs that omit required functionality or fail to address the requested programming task. We randomly sample 15 problems and manually inspect outputs from each model across all prompt variants. The same 15 tasks are evaluated across all conditions to ensure fair comparison. The sample size was selected to balance coverage with the manual effort required for detailed inspection. For each output, we check whether (i) the solution addresses the stated programming task, (ii) required functionality is implemented, and (iii) no degenerate responses (e.g., empty stubs, constant-return programs, or unrelated templates) are produced. All artifacts supporting this evaluation are available online ~\cite{anonymous2026empirical}.

\FloatBarrier

\section{Results}

\subsection{Stage 1 (RQ1): Prompt-Level Generation Compliance and Security Weakness Prevalence}
We first analyze prompt-level effects across all prompt variants (Prompt 0.0--4.0), focusing on (i) the prevalence of invalid or non-code outputs and (ii) the number of generated files flagged as vulnerable by Bandit. This analysis quantifies how prompt structure affects refusal behaviour and the prevalence of security weaknesses in generated code.

\begin{figure}[ht!]
    \centering
    \includegraphics[width=1\columnwidth]{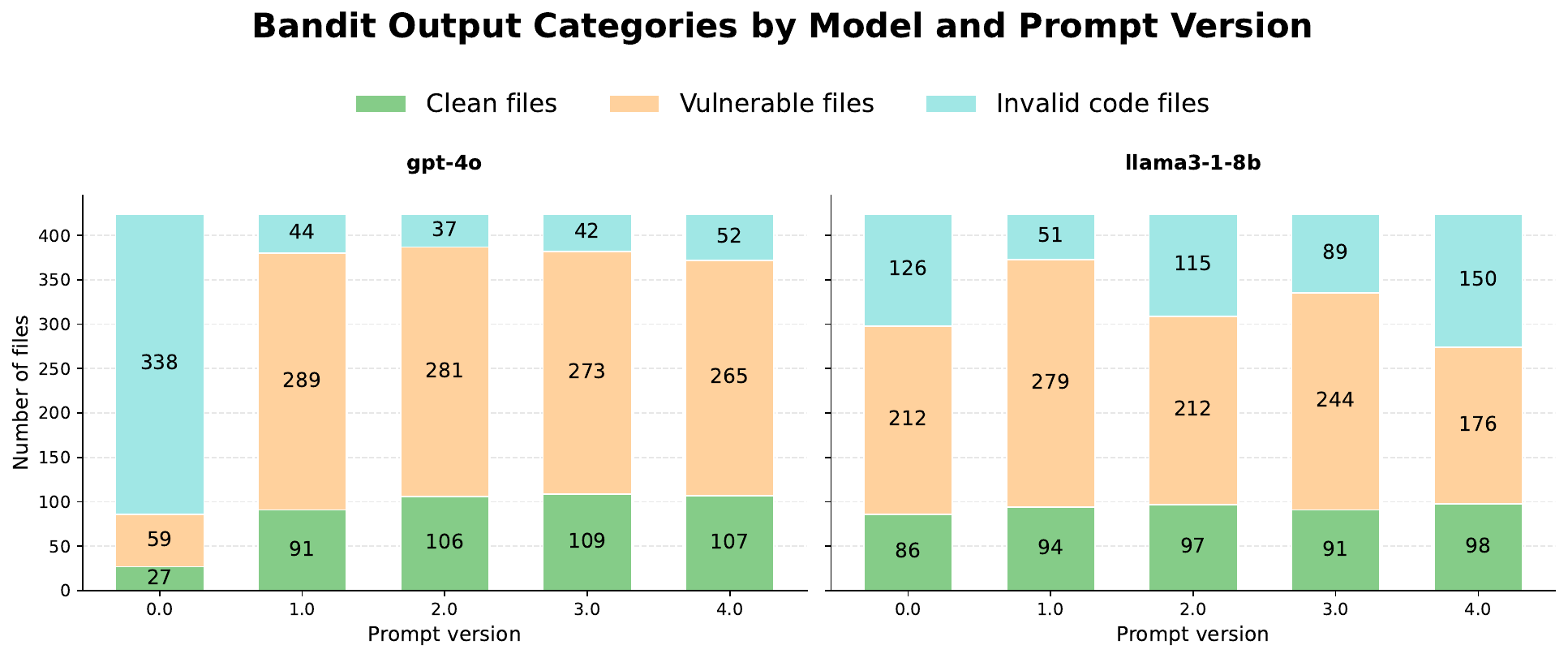}
    \caption{Bandit analysis outcomes across prompt variants for GPT-4o and LLaMA 3.1-8B (424 tasks per prompt).}
    \label{fig:bandit_prompt_comparison}
\end{figure}

Figure~\ref{fig:bandit_prompt_comparison} summarizes the results for GPT-4o and LLaMA 3.1-8B. 
Under the minimal Prompt 0.0 baseline, both models exhibit high refusal rates, though the effect is substantially more pronounced for GPT-4o. Of the 424 tasks, GPT-4o produces only 86 syntactically valid Python files, of which 59 are flagged as vulnerable, and 27 contain no Bandit findings. The remaining 338 outputs are classified as invalid or non-code responses. This indicates that, in the absence of explicit structural guidance, GPT-4o frequently declines to generate code for security-sensitive tasks. 

Introducing structured prompting in Prompt 1.0 leads to a sharp shift in generation behaviour for GPT-4o. The number of invalid outputs decreases from 338 to 44, while the number of valid Python files increases accordingly. This improvement remains stable across Prompts 2.0 through 4.0, with invalid outputs consistently ranging between 37 and 52 files. These results suggest that explicit output constraints, formatting requirements, and layered instructions are sufficient to substantially reduce refusal behaviour. The observed pattern indicates that the absence of structured prompting, rather than task complexity alone, is a primary contributor to refusals under Prompt 0.0.

LLaMA 3.1-8B exhibits a markedly different response. While Prompt 1.0 reduces invalid outputs from 126 to 51, subsequent prompt refinements do not yield consistent gains. Invalid outputs increase again in Prompts 2.0-4.0 (115, 89, and 150 files, respectively), indicating that additional prompt structure has limited and unstable effects on compliance for this model. This contrast highlights that the compliance benefits of structured prompting may be strongly model-dependent.

We next examine the number of valid Python files flagged as vulnerable by Bandit. For GPT-4o, security weakness counts decrease modestly across structured prompts, from 289 in Prompt 1.0 to 265 in Prompt 4.0. While this trend suggests that security-oriented prompt refinements may influence security issue prevalence, the reduction is small relative to the dramatic gains observed in code generation compliance.

For LLaMA 3.1-8B, no clear monotonic trend is observed. Vulnerable file counts fluctuate across prompt versions (279 in Prompt 1.0, 212 in Prompt 2.0, 244 in Prompt 3.0, and 176 in Prompt 4.0), indicating that prompt refinement does not consistently reduce the prevalence of security weaknesses for this model.


\subsection{Stage 2 (RQ2): Security Weakness Severity and Type Analysis}

Stage 2 characterizes security weakness \emph{severity} and \emph{type} in LLM-generated code, conditioning on successful code generation. Since invalid and non-code outputs were considered in Stage 1, we filter out this subset to enable a fair analysis of only syntactically valid Python code files. The goal of this stage is to determine whether structured, security-oriented prompting alters not only the \emph{presence} of security-relevant issues, but their \emph{severity profile} and \emph{underlying security weakness classes}. 

We proceed in two steps. First, we analyze GPT-4o in isolation because its higher code-generation compliance yields a larger and more stable set of analyzable samples for fine-grained security weakness analysis. We then perform the same severity and CWE analyses on the filtered LLaMA 3.1-8B outputs, retaining only tasks that produced syntactically valid Python code under all evaluated prompt conditions. This enables us to observe how the severity of security issues and the distributions of CWEs evolve under prompt refinement for each model.

\subsubsection{Severity Distribution Across Prompt Variants}

Figure~\ref{fig:bandit_severity_comparison} reports Bandit severity distributions for both models across all structured prompt variants. We first focus on GPT-4o to isolate within-model prompt effects. For GPT-4o, the proportion of high-severity findings decreases from 20.8\% under Prompt 1.0 to 13.6\% under Prompt 4.0. Along the same progression, the proportion of low-severity findings increases from 32.0\% to 43.5\%, while the proportion of medium-severity findings decreases modestly from 47.1\% to 43.0\%. Most of the reduction in high-severity findings occurs immediately after the introduction of explicit security guidance in Prompt 2.0. Subsequent prompt refinements produce comparatively smaller changes, suggesting that the presence of security-oriented instructions may be more influential than the specific form those instructions take.

These results suggest that increasingly security-oriented prompts shift the severity profile toward lower-severity findings rather than eliminating security weaknesses outright. Importantly, Bandit severity levels are not uniformly distributed across security weakness classes. HIGH-severity findings generally correspond to insecure patterns that Static Application Security Testing (SAST) tools associate with greater security risk, whereas low-severity findings are often associated with contextual weaknesses and coding practices, such as insufficient exception handling or information exposure. Consequently, the observed shift in severity reflects changes in the types of security weaknesses being reported rather than a simple redistribution among severity categories.

In contrast, LLaMA 3.1-8B exhibits no comparable monotonic trend. The proportion of HIGH-severity findings remains relatively stable across prompt variants (21.2\%, 22.9\%, 21.5\%, and 24.4\% for Prompts 1.0–4.0, respectively), indicating that prompt refinement has a weaker effect on severity outcomes for this model. This divergence suggests that prompt-induced severity shifts may vary across the evaluated models. While GPT-4o exhibits a noticeable redistribution of security issue severity across prompt variants, LLaMA 3.1-8B maintains comparatively stable severity distributions. These results indicate that the security effects of prompt refinement are not uniform across the models examined in this study.

\subsubsection{CWE-Level Security Weakness Analysis}

\begin{figure}[t]
    \centering
    \includegraphics[width=\columnwidth]{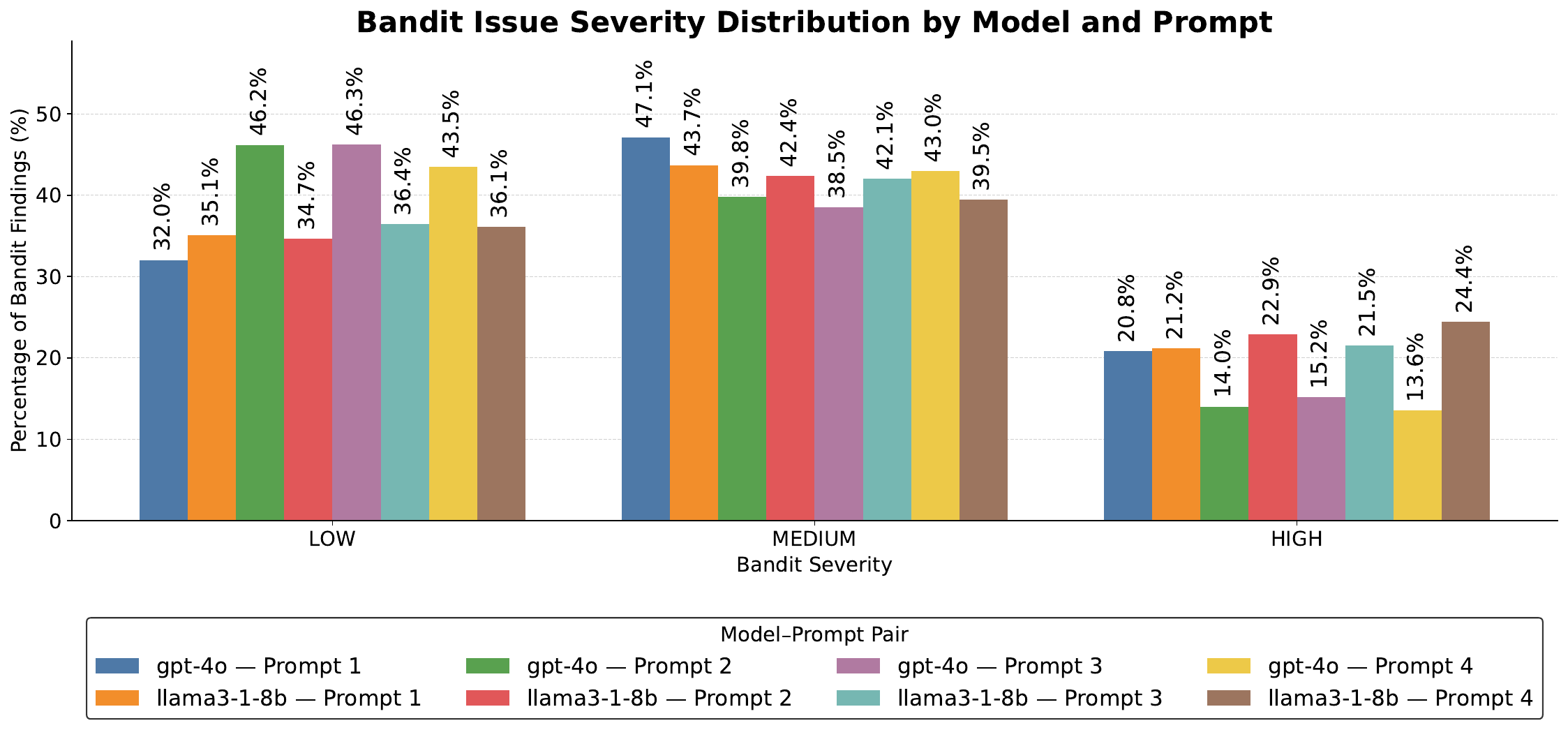}
    \caption{Bandit severity distribution across prompt variants.}
    \label{fig:bandit_severity_comparison}
\end{figure}
CWE identifiers follow the Common Weakness Enumeration taxonomy~\cite{mitre_cwe}. Table~\ref{tab:cwe_prompt_distribution} reports the top 10 most frequent CWE identifiers detected by Bandit and CodeQL across all prompt variants and both models. These CWEs account for the majority of observed security weaknesses and capture the dominant security trends in the dataset. A key observation is that the most prevalent security weakness classes remain dominant across all prompt variants. CWE-78 (OS Command Injection) and CWE-502 (Deserialization of Untrusted Data) consistently account for a large proportion of detected findings regardless of prompt structure. For GPT-4o, CWE-78 increases from 43.9\% under Prompt 1.0 to 49.7\% under Prompt 4.0, while CWE-502 decreases modestly from 23.7\% to 21.0\%. Although the magnitude and direction of change vary across prompt variants, both CWE categories remain highly prevalent throughout the evaluation. The persistence of CWE-78 and CWE-502 suggests that prompt refinement does not affect all security weakness classes equally. While some weaknesses become less frequent under increasingly security-oriented prompts, these dominant security weakness categories remain common across prompt conditions, indicating that prompt-level interventions may have limited influence on certain classes of security-relevant issues.

In contrast, security weakness classes associated with explicit dynamic code execution show greater prompt sensitivity for GPT-4o. CWE-94 (Improper Control of Generation of Code) decreases from 10.1\% under Prompt 1.0 to 4.1\% under Prompt 4.0, although the trend is not strictly monotonic. CWE-95 (Improper Neutralization of Directives in Dynamically Evaluated Code) and CWE-116 (Improper Encoding or Escaping of Output) exhibit similar reductions, decreasing from 3.1\% under Prompt 1.0 to 0.6\% under Prompt 4.0. Together, these patterns are consistent with GPT-4o adopting implementation strategies that avoid dynamic code execution patterns such as \texttt{eval()} and \texttt{exec()}.

For LLaMA 3.1-8B, these same CWE classes remain comparatively stable. CWE-94 varies between 8.6\% and 10.2\% across prompt variants and exhibits no consistent downward trend, while CWE-95 and CWE-116 change only modestly. This contrast helps explain the absence of a monotonic severity reduction for LLaMA 3.1-8B in Figure~\ref{fig:bandit_severity_comparison} and highlights the model-dependent nature of prompt effectiveness.

\definecolor{deltacol}{gray}{0.92}
\definecolor{headergray}{gray}{0.88}

\begin{table}[htbp]
\centering
\scriptsize
\caption{Security weakness (SW) prevalence and top 10 CWE distribution across prompts. The $\Delta$ column reports the percentage-point change from P1.0 to P4.0, where positive values indicate an increase and negative values indicate a decrease.}
\label{tab:cwe_prompt_distribution}
\begin{adjustbox}{max width=\textwidth}
\setlength{\tabcolsep}{3pt}
\begin{tabular}{lccccc|ccccc}
\toprule
\textbf{Measure} &
\multicolumn{5}{c|}{\textbf{GPT-4o}} &
\multicolumn{5}{c}{\textbf{LLaMA 3.1-8B}} \\
\cmidrule(lr){2-6} \cmidrule(lr){7-11}
& \textbf{P1.0} 
& \textbf{P2.0} 
& \textbf{P3.0} 
& \textbf{P4.0} 
& \cellcolor{headergray}\textbf{$\Delta$}
& \textbf{P1.0} 
& \textbf{P2.0} 
& \textbf{P3.0} 
& \textbf{P4.0} 
& \cellcolor{headergray}\textbf{$\Delta$} \\
\midrule

\rowcolor{gray!12}
\textbf{\begin{tabular}[c]{@{}c@{}}
(\%) SW\\
prevalence 
\end{tabular}}
& \textbf{76.3} 
& \textbf{72.6} 
& \textbf{72.5} 
& \textbf{72.0} 
& \cellcolor{deltacol}\textbf{-4.3}
& \textbf{75.0} 
& \textbf{69.6} 
& \textbf{72.8} 
& \textbf{65.0} 
& \cellcolor{deltacol}\textbf{-10.0} \\

\midrule
\multicolumn{11}{l}{\textit{CWE (\%)}} \\
\midrule

CWE-78  
& 43.9 & 49.2 & 49.1 & 49.7 & \cellcolor{deltacol}\textbf{+5.8}
& 43.3 & 38.8 & 41.4 & 38.8 & \cellcolor{deltacol}\textbf{-4.5} \\

CWE-502 
& 23.7 & 23.3 & 22.7 & 21.0 & \cellcolor{deltacol}\textbf{-2.7}
& 26.1 & 25.0 & 25.4 & 21.4 & \cellcolor{deltacol}\textbf{-4.7} \\

CWE-94  
& 10.1  & 4.8  & 5.9  & 4.1  & \cellcolor{deltacol}\textbf{-6.0}
& 9.3  & 8.6  & 9.9  & 10.2  & \cellcolor{deltacol}\textbf{+0.9} \\

CWE-209 
& 4.2  & 3.6  & 3.7  & 4.5  & \cellcolor{deltacol}\textbf{+0.3}
& 5.1  & 4.5  & 4.2  & 5.6  & \cellcolor{deltacol}\textbf{+0.5} \\

CWE-497 
& 4.2  & 3.6  & 3.7  & 4.5  & \cellcolor{deltacol}\textbf{+0.3}
& 5.1  & 4.5  & 4.2  & 5.6  & \cellcolor{deltacol}\textbf{+0.5} \\

CWE-489 
& 0.9  & 2.9  & 3.5  & 2.1  & \cellcolor{deltacol}\textbf{+1.2}
& 0.2  & 3.0  & 2.1  & 2.9  & \cellcolor{deltacol}\textbf{+2.7} \\

CWE-215 
& 0.9  & 2.9  & 3.5  & 2.1  & \cellcolor{deltacol}\textbf{+1.2}
& 0.2  & 3.0  & 2.1  & 2.9  & \cellcolor{deltacol}\textbf{+2.7} \\

CWE-116 
& 3.1  & 0.5  & 0.7  & 0.6  & \cellcolor{deltacol}\textbf{-2.5}
& 2.9  & 2.2  & 2.9  & 2.7  & \cellcolor{deltacol}\textbf{-0.2} \\

CWE-95  
& 3.1  & 0.5  & 0.7  & 0.6  & \cellcolor{deltacol}\textbf{-2.5}
& 2.9  & 2.2  & 2.9  & 2.7  & \cellcolor{deltacol}\textbf{-0.2} \\

CWE-330 
& 1.3  & 1.2  & 1.3  & 1.3  & \cellcolor{deltacol}\textbf{0.0}
& 1.7  & 2.2  & 1.5  & 1.7  & \cellcolor{deltacol}\textbf{0.0} \\

\bottomrule
\end{tabular}
\end{adjustbox}

\vspace{0.5em}
\footnotesize
\textit{Note.} Security weakness prevalence is computed as the percentage of valid Python files flagged by at least one SAST tool. CWE percentages are computed over cumulative CWE findings for each model--prompt pair.
\end{table}

Several CWE classes related to information exposure and debugging behaviour, including CWE-209 (Generation of Error Message Containing Sensitive Information), CWE-497 (Exposure of Sensitive System Information to an Unauthorized Control Sphere), CWE-489 (Active Debug Code), and CWE-215 (Insertion of Sensitive Information Into Debugging Code), exhibit non-monotonic trends across prompt variants. However, the direction and magnitude of change differ across security weakness classes. CWE-209 and CWE-497 fluctuate within a relatively narrow range, whereas CWE-489 and CWE-215 increase under some intermediate prompt conditions before partially declining. These irregular patterns suggest that prompt refinement does not exert a uniform influence on information-exposure and debugging-related weaknesses. Finally, CWE-330 (Use of Insufficiently Random Values) remains effectively unchanged across prompt variants for both models.

Overall, Stage 2 demonstrates that structured, security-oriented prompting can alter the severity profile and security-relevant issue composition of LLM-generated code, although these effects vary across models and security weakness classes. For GPT-4o, prompt refinement is associated with a lower proportion of HIGH-severity findings and reductions in several dynamic-execution-related CWE categories. In contrast, LLaMA 3.1-8B exhibits comparatively stable severity and CWE distributions across prompt variants. At the same time, several HIGH-frequency security weakness classes, including CWE-78 and CWE-502, remain prevalent regardless of prompt structure, suggesting that some security weaknesses are substantially less responsive to prompt refinement than others. These findings indicate that prompt engineering can influence the distribution of security risk, but should be viewed as a complementary measure rather than a substitute for dedicated security analysis and post-generation validation.

Taken together, the CWE results reveal an uneven response to prompt refinement. The largest decreases are concentrated in a relatively small subset of security weakness categories, particularly CWE-94, CWE-95, and CWE-116 for GPT-4o, while the most prevalent security issue classes remain common across all prompt variants. This pattern suggests that prompt refinement influences the composition and manifestation of security weaknesses more strongly than it influences their overall prevalence.

\subsection{Task Relevance and Security-Driven Semantic Drift}

We conducted a task-relevance analysis on a random sample of 15 problem IDs. For each sampled task, we inspected solutions generated by both models across all structured prompt variants (Prompts 1.0-4.0) to determine whether outputs remained contextually aligned with the original coding requirements.
Across all sampled tasks and prompt variants, all generated outputs were found to be task-relevant. No outputs were classified as unrelated or nonsensical, indicating that models did not bypass security-sensitive tasks by emitting trivial or placeholder code.
This suggests that the observed security weakness trends in earlier analyses are not artifacts of task abandonment but rather reflect substantive changes in how tasks are implemented.

Beyond task relevance, however, we observe a recurring pattern of
\emph{security-driven semantic drift}, particularly under more structured and security-oriented prompts. In multiple cases, models mitigated security weaknesses by replacing or removing explicitly requested unsafe constructs with safer alternatives,
such as substituting \texttt{eval()} with AST-based evaluation, preferring \texttt{subprocess} over \texttt{os.system}, or introducing stricter input validation. While these transformations often reduced detectable security weaknesses, they sometimes altered the prescribed implementation strategy, replacing explicitly required unsafe constructs with safer alternatives while still addressing the core task objective.

This phenomenon is observed in 20\% of tasks under the baseline structured prompt (Prompt 1.0), but becomes increasingly prevalent under security-guided and adversarial prompts (Prompts 2.0-4.0), where models more aggressively rewrite or suppress dangerous
primitives. Table~\ref{tab:semantic_drift} quantifies this effect across the sampled tasks. For GPT-4o, semantic drift rises sharply from 20\% under Prompt 1.0 to 60-67\% under Prompts~2.0--4.0, indicating a strong tendency to prioritize security compliance over
adherence to task-specific API requirements as prompt constraints intensify. Drift probability increases 3× under security-guided prompts. In many of these cases, unsafe constructs central to the task specification are replaced or removed entirely, despite being explicitly required.

The increase in semantic drift is substantially larger than the observed reduction in security weakness prevalence. This suggests that security-oriented prompting frequently changes implementation strategies even when security-relevant issues remain present, reinforcing the view that prompt refinement primarily reshapes risk rather than eliminating it.

\begin{table}[htbp]
\centering
\setlength{\tabcolsep}{4pt}
\renewcommand{\arraystretch}{1.15}
\caption{Incidence of security-driven semantic drift across prompt variants. 
Drift denotes relevant outputs that alter or remove explicitly required unsafe constructs.}
\label{tab:semantic_drift}
\begin{tabular}{lcccc}
\hline
\textbf{Model} & \textbf{P1.0} & \textbf{P2.0} & \textbf{P3.0} & \textbf{P4.0} \\
\hline
GPT-4o & 3/15 (20\%) & 9/15 (60\%) & 10/15 (67\%) & 9/15 (60\%) \\
LLaMA~3.1~8B & 1/15 (7\%) & 3/15 (20\%) & 3/15 (20\%) & 4/15 (27\%) \\
\hline
\end{tabular}
\end{table}

In contrast, LLaMA 3.1-8B exhibits substantially lower drift rates across all prompt variants. Semantic drift increases more gradually from 7\% under Prompt~1.0 to 20-27\%
under Prompts 2.0--4.0, suggesting a more conservative response to security guidance. While this behaviour results in fewer consistent reductions in detected security weaknesses, it more reliably preserves exact API usage and task semantics across prompt variants.

These findings highlight an important trade-off introduced by security-oriented prompt refinement. While structured prompts can encourage safer implementations and reduce static-analysis findings, they may also induce semantic drift that departs from explicit task requirements. Importantly, our relevance audit found that sampled outputs remained aligned with the original programming objectives, indicating that the observed reductions in security weakness severity are not simply explained by task abandonment or the generation of unrelated code. Instead, security-oriented prompts frequently preserved the underlying functionality while altering implementation details, often replacing explicitly requested unsafe constructs with safer alternatives. This suggests that prompt refinement primarily influences how security-sensitive functionality is implemented rather than whether it is implemented, shifting generated code toward safer approaches without necessarily eliminating the underlying security risk.

\subsection{Qualitative Analysis of Unflagged Security-Relevant Outputs}

To better understand the limitations of static-analysis-based evaluation, we manually reviewed outputs that were not flagged by either Bandit or CodeQL. Although these samples were considered free of detectable weaknesses by both tools, manual inspection revealed several recurring security-relevant patterns.

A common theme was the presence of security weaknesses that depended on program semantics, runtime behaviour, or application context, rather than the easily pattern-matched weaknesses typically targeted by static-analysis rules. Examples included user-controlled SQL query execution, timing side channels arising from password-comparison logic, and interactive Python consoles that effectively exposed arbitrary code-execution capabilities without relying on commonly flagged functions such as \texttt{eval()} or \texttt{exec()}. We also observed multiple resource-exhaustion risks, including excessive memory allocation, deep recursion, infinite loops, and large-scale thread creation. In several cases, generated programs exposed potentially sensitive system functionality through command-dispatch or information-retrieval interfaces despite avoiding APIs traditionally associated with command injection.

These observations do not establish that static-analysis tools failed to detect confirmed weaknesses in every case. Rather, they illustrate classes of security-relevant behaviour that may not be captured by pattern-based detection rules. Consequently, security weakness prevalence derived from Bandit and CodeQL should be interpreted as a lower-bound estimate of security risk. While both tools remain effective for identifying many common weakness patterns, security-relevant behaviours that emerge from program logic, resource consumption, or application-specific context may require complementary forms of analysis beyond conventional static scanning.

\section{Discussion}

Beyond evaluating whether prompting reduces security weaknesses, this study characterizes how different forms of prompt guidance influence security-related behaviour in code generation. Across the evaluated models, prompt refinements consistently affected compliance and severity distributions, but their effects on overall prevalence of security weaknesses were comparatively limited. Structured prompting therefore functions primarily as a compliance and behavioural shaping mechanism rather than a direct security fix. For GPT-4o, adding structure dramatically reduced refusals and enabled large-scale security analysis. However, this introduces a caveat in the evaluation: prompts that reduce refusals change the denominator of analyzable outputs. Comparing counts of security weaknesses without conditioning on successful code generation can therefore misattribute improvements to security rather than compliance.

Conditioned on valid Python code, security-oriented prompts did not consistently reduce security weakness prevalence. Structured prompting reduces the proportion of high-severity findings, but produces only modest and inconsistent changes in overall weakness prevalence, suggesting redistribution rather than elimination as the dominant effect. For GPT-4o, stronger prompts shifted the severity distribution (lowering high-severity scores and raising low-severity scores), whereas LLaMA exhibited weaker or inconsistent effects. These findings suggest that prompt structure can influence implementation choices (e.g., avoiding dynamic execution patterns) but cannot reliably mitigate insecure implementations embedded in task semantics (e.g., command execution or deserialization). Moreover, semantic drift reveals that security-oriented prompting can modify implementation strategies without eliminating security-relevant issues, introducing a trade-off between security compliance and fidelity to the original task specification: stronger prompts often replace explicitly requested unsafe constructs with safer alternatives, reducing static findings while altering intended implementations. Overall, prompt engineering shapes compliance and risk distribution, but remains insufficient as a standalone security control. Critically, evaluations that rely on aggregate counts of security weaknesses may overstate the security benefit of prompt refinement, because a reduction in high-severity findings accompanied by an increase in low-severity findings appears as an improvement under count-based metrics while leaving total presence of security weakness unchanged. This suggests the field needs severity-stratified evaluation metrics as a baseline standard for assessing prompt-based security interventions. For practitioners, these findings suggest that security-oriented prompts should be treated as a complementary aid rather than a replacement for established security practices. While prompt refinement can influence security weakness severity and composition, static analysis, code review, and security testing remain necessary components of secure AI-assisted development.

\textbf{Threats to Validity. }First, we rely on static analysis (Bandit and CodeQL), which may produce false positives and fail to detect certain vulnerabilities~\cite{ami2023falsenegative}. Static analysis tools approximate security risk through pattern- and data-flow-based reasoning and do not directly assess runtime behaviour or exploitability. As a result, some reported weaknesses may not be exploitable in practice, while others may remain undetected. Second, our evaluation is limited to Python tasks from a single vulnerability-focused dataset, which may restrict generalizability to other languages, domains, or development contexts. Third, prompt effects are model-dependent. GPT-4o and LLaMA 3.1-8B responded differently to structured prompting, and results may not generalize to other architectures or future model versions. Additionally, we generate a single output per configuration, limiting variance estimation. Future work should incorporate dynamic testing, multi-sample evaluation, additional languages, and broader prompt strategies.

Finally, the Cybernative.ai dataset consists of synthetic security-oriented tasks. Manual inspection indicated that some prompts resemble realistic developer requests, whereas others explicitly encourage insecure coding practices. As a result, the observed security weakness distributions may be influenced by the benchmark's security-focused nature.

\section{Conclusion}
We presented an empirical study on the use of prompt structure as a security intervention for LLM-generated code. Across 424 security-sensitive Python tasks, structured prompts largely eliminated refusals and enabled security weakness assessment at scale, but did not consistently reduce the overall prevalence of security weaknesses. Instead, effects were model-dependent: for GPT-4o, stronger security prompts primarily shifted findings from high- to low-severity and often induced security-driven semantic drift, whereas LLaMA showed weaker, less consistent changes. Overall, prompt engineering is a strong compliance lever, but it should complement, not replace, post-generation security controls.

Several directions follow from these findings. The study could be broadened beyond two models, a single Python dataset, and zero-shot prompting, extending to additional model families, languages, and prompt strategies such as few-shot and chain-of-thought reasoning that vary a second design dimension held fixed here. Finally, our semantic-drift audit could be scaled and coupled with functional test execution to measure how often security-driven rewrites preserve the originally specified behaviour, clarifying the trade-off between security weakness reduction and task fidelity.

\section*{Acknowledgements}
OpenAI ChatGPT and Grammarly were used to assist with language editing and manuscript refinement. All experiments, analysis, and conclusions were verified by the authors.

\bibliographystyle{IEEEtran}
\bibliography{main.bib}
\end{document}